# Novel polymer nanocomposite composed of organic nanoparticles *via* self-assembly

*Dequan Xiao,[1][*] Kunhua Lin,[2] Qiang Fu[3], Qinjian Yin,[2][*]*

[1]Department of Chemistry, Duke University
Durham, NC 27705, USA
E-mail: dequan.xiao@duke.edu

[2]Department of Chemistry, Sichuan University
Chengdu, Sichuan, 610064, China
Email: Changer@scu.edu.cn

[3]Department of Polymer Science & Engineering
Sichuan University
State Key Laboratory of Polymer Materials Engineering
Chengdu, Sichuan, 610065, China

**Abstract**

We report a novel class of polymer nanocomposite composed of organic nanoparticles dispersed in polymer matrix, with the particle sizes of 30-120 nm in radius. The organic nanoparticles were formed by the self-assembly of protonated poly(4-vinyl-pyridine)-*r*-poly(acrylonitrile) and amphiphilic metanil yellow dye molecules through electrostatic interactions in aqueous solution. A strongly broadened Raman shift band was probed, suggesting the presence of enhanced optoelectronic property from the polymer nanocomposite. Here, using random-copolymer polyelectrolytes and mesogenic amphiphiles as the designed building blocks for self-assembly, a new approach is acutally provided to fabricate organic nanoparticles.

**1. Introduction**

Designing and engineering effective molecular or polymer building blocks [1-3] for self-assembly [4-6] is one of the challenges toward the pathway to fabricate novel organic nanostructured materials with particular morphology and functions [1]. Here, using well-designed building blocks of a random-copolymer polyelectrolyte and amphiphilic mesogenic azobenene dye molecules for self-assembly, we report a novel class of polymer nanocomposite composed of organic nanoparticles from which a strongly broadened Raman shift band was probed.

By the building blocks of copolymers, such as block copolymers and random copolymers, different functional monomer segments can be incorporated into one polymer chain, providing the structural feature for the formation of abundant nano-phases in condensed states. In the last 30 years, block copolymers [7, 8] have been employed intensively as extremely effective building blocks to achieve fruitful types of nanostructures such as lamellae, cylinders, gyroid, spheres and as well as hierarchical nanostructures in a controllable means. For this purpose, different types of block copolymers such as crystalline-amorphous [9-14] and liquid crystalline [10, 15-19] block copolymers have explored. For applications, block copolymer lithography has been developed as a promising technique to form nano-patterns for manufacturing integrated circuits [20, 21]. In recent years, attention has been caught on building self-assembled nanostructures using the building blocks of block copolymers and amphiphilic small (low molecular weight) organic molecules through non-covalent interactions such as electrostatic interactions [22-25] and hydrogen bonding [26-31]. For instances, hierarchical nanostructues were obtained using the self-assembly of polystyrene-*b*-poly(4-vinyl-pyridine) (PS-*b*-P4VP) and amphiphilic bent-core molecules [27] through hydrogen bonding; supramolecular cylinder structures were generated by the self-assembly of protonated P4VP one-block polymer and amphiphilic azobenene derivatives through electrostatic interactions [25]; and using PS-*b*-P4VP and nonadecylphenol (NDP), hierarchical nanostructures of spherical (PS) in lamellar (P4VP(NDP)$_{1.0}$) matrix was obtained through hydrogen bonding, with the sphere sizes less than 50 nm in radius. However, less spherical nanostructures containing small organic molecules (in either pure or composite form) *via* the self-assembly of block copolymers and small organic moelcuels has been reported. Compared to block copolymers, random copolymers have not been used extensively as self-assembly building blocks to fabricate well-defined nanostructures because less control on polymer chain structures can be conducted for random copolymers than for block copolymers. However, through a gradual hydrophobic aggregation mechanism in micelles, an amphiphlic random copolymer containing hydrophobic azobenene side chains was used to form colloidal spheres with the radius of 40-90 nm by taking advantage of the polarity difference among monomer segments [32].

Spherical organic nanostructures composed by aggregated small organic moelcules (also known as organic nanoparticles) have recently attracted considerable attention [33] because of their special optoelectronic properties related to the geometry and morphology of the nanostructures [34, 35]. Pure organic nanoparticles of π-conjugated small organic molecules such as perylene, phthalocyanine, pyrazoline and perylene diimde have been prepared by reprecipitation (solvent displacement) method [36, 37], effective colloid chemical reaction method [38] and self-assembly of π-conjugated molecules with steric-hindered swallow-tail-like side chains [39]. In addition, composite organic nanoparticles (nanoscale mixtures of organic small molecules and polymers) have been prepared through a mechanism of polymer microemulsion micelles [40]. However, the preparation of organic nanoparticles dispersed in polymer matrix using the self-assembly of random-copolymer polyelectrolyte and amphiphlic small molecules has not been reported.

The hybrid materials with organic nanoparticles dispersed in polymer materix can be called as a novel class of polymer nanocomposite [41-43]. Using polymer nanocomposite such as the materials with added inorganic nanoparticles in polymers, improved property performance has been achieved [32, 44, 45]. Analogously, fabricating polymer nanocomposite containing organic nanoparticles using azobezene dyes can lead to enhanced optoelectronics properties for the applications of photoswitching, optical data-storage, and sensors. For this reason, photoresponsive uniform colloidal spheres at nanoscale in water suspensions have been prepared from amphiphilic azobenzene-containing random polymer [32].

In this communication, two types of building blocks: protonated poly(4-vinyl-pyridine)-*r*-poly(acrylonitrile) (P4VP-*r*-PAN) random copolymer and anionic metanil yellow (MY) dye molecules (3-[[4-(phenylamino) phenyl]azo]-benzene sulfonic acid monosodium salt) were employed to perform self-assembly. The two building blocks interact with each other through electrostatic interactions in aqueous solution to obtain spherical organic nanostructures. For comparison, the morphology of pure chloride salt of the P4VP-*r*-PAN polyelectrolyte was characterized by scanning electron microscopy (SEM) and transmission electron microscopy (TEM). Morphologies of the organic nanoparticles dispersed in polymer matrix were characterized by transmission electron microscopy (TEM). A strongly broadened Raman shift band was probed for the polymer nanocomposite, compared to the Raman shift spectrum of natural MY dye.

## 2. Experimental section

### 2.1. Preparation of organic nanoparticles dispersed in polymer matrix

The organic nanoparticles were prepared *via* the self-assembly of protonated P4VP-*r*-PAN and MY molecules, and the chemical structures of typical P4VP-*r*-PAN polymer chain bound by MY molecules is schematically shown in Figure 1. The P4VP-*r*-PAN copolymer was synthesized by radical polymerization [46], neutralized by HCl acid, and then prepared into an aqueous solution of $1.0 \times 10^{-3}$ M (based on 4-vinylpyridine repeat units). A aqueous solution of MY dye (Aldrich) in stoichiometically equalent concentration ($1.0 \times 10^{-3}$ M) was also prepared. The P4VP-*r*-PAN and MY solutions were mixed to induce phase seperation. The precipated polymer nanocomposite was filtered out, washed with copious amounts of water, and dried under vacuum at 60 °C for 2 days. From differential scanning calorimetry thermal diagram, the dried polymer nanocomposite show no phase transition until 193ºC, indicating that the organic nanoparticles have high thermal stability [46]. The complexation of P4VP-*r*-

PAN and MY by electrostatic interactions between the pyridinum and sulfonic groups were supported by the evidences of FT-IR spectra [46].

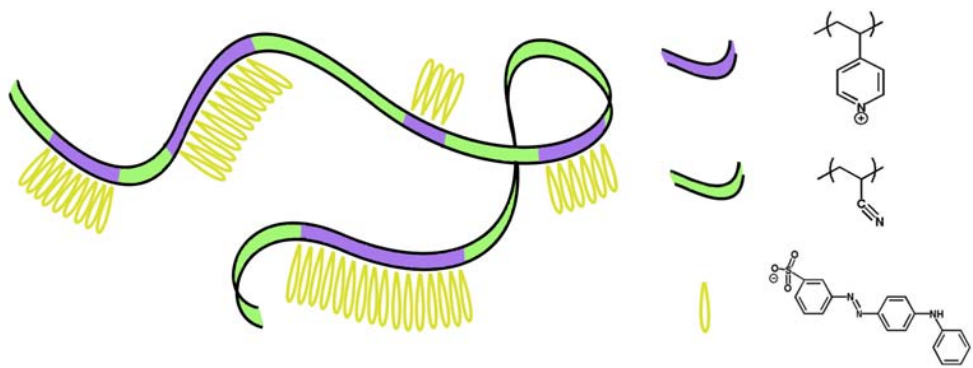

**Figure 1**. Schematic representation of a typical cationic polymer chain of protonated P4VP-*r*-PAN bound by anionic metanil yellow dye molecules. The pyridinium-containing monomer segments are in purple, the acrylonitrile-containing monomer segments are in green, and the dye molecules are represented in yellow.

**2.2. Characterization and analysis**

The TEM micrographs of the P4VP-*r*-PAN polyelectrolyte and the polymer nanocomposite containing organic nanoparticles were obtained by JEOL-100CX TEM from JEOL Ltd. (Japan). The SEM micrographs were obtained by Hitachi s-4800 from Hitachi Ltd. (Japan). The samples for TEM and SEM were vacuum-dried for 24 hours before measurement.

For Raman shift spectra, the solid samples of natural MY and the polymer nanocomposite containing organic nanoparticles were respectively analyzed on an InVia Raman microscope from Renishaw (United Kingdom). The samples were excited by a 632 nm laser with 30 s integration time. For comparison, Raman activity spectrum of the MY acid form was calculated by Gaussian program packages [47] using Hartee-Fock method in 3-21G* basis set, while the geometry optimization was performed using Hartee-Fock method in 6-31G* basis set. The calculated Raman activity spectrum was adjusted by a factor of 0.89 and was normalized based on the strongest experimental peak to compare with the experimental Raman shift spectrum.

## 3. Result and Discussion

### 3.1. Morphologies of the P4VP-*r*-PAN copolymer and the MY organic nanoparticles

Figure 2 shows the morphologies of pure chloride salt of the protonated P4VP-*r*-PAN in bulk materials that were formed after evaperization of water. The P4VP-*r*-PAN polyelectrolyte surface (Figure 2a) is generally smooth, however, hollow spaces (cracks) are observed on the surface at the nanoscale. From the TEM micrograph (Figure 2b) , the surface cracks can be correlated to the porous 3D network structure of the polyelectrolyte, where the hollow spaces are formed by entangled polymer chains. The porous 3D network structure of the polyelectrolyte are also found in the polymer nanocomposite containing organic nanoparticles as shown below.

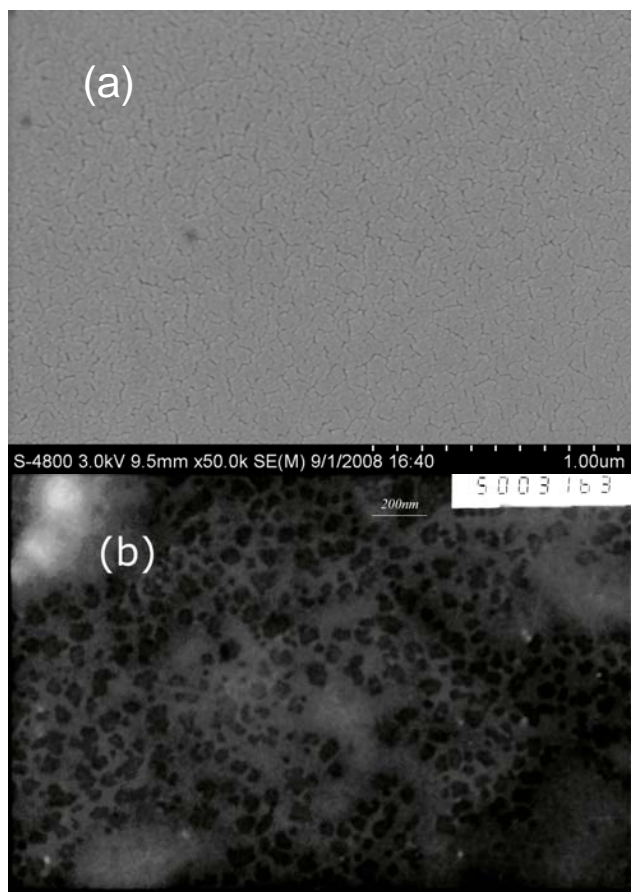

**Figure 2.** (a) SEM morphology and (b) TEM morphology of dried chloride salt of the P4VP-*r*-PAN polyelectrolyte.

Figure 3 shows the morphologies of organic nanoparticles dipersed in polymer matrix. The organic nanoparticles are in spherical shape with the sizes of 30-120 nm in radius (see Figure 4), and the entangled polymer chains are shown as the porous network structure similar to that in Figure 2b. With the change of the initial concentration of 4VP units or MY molecules in aqueous solution (from $1.0\times10^{-3}$ M to $1.0\times10^{-5}$ M in Figure 3a-b) for the preparation procedure, the formed organic nanoparticles remain in spherical shape and have shown no apparent change in size distributions. From the viewpoint of micellization mechanism, the organic nanoparticles are likely formed as the micelles with MY dye molecules in the core and P4VP-*r*-PAN copolymer in the corona. However, the exact chemical composition and molecular arrangments of the organic nanoparticles need to be elucidated in the future.

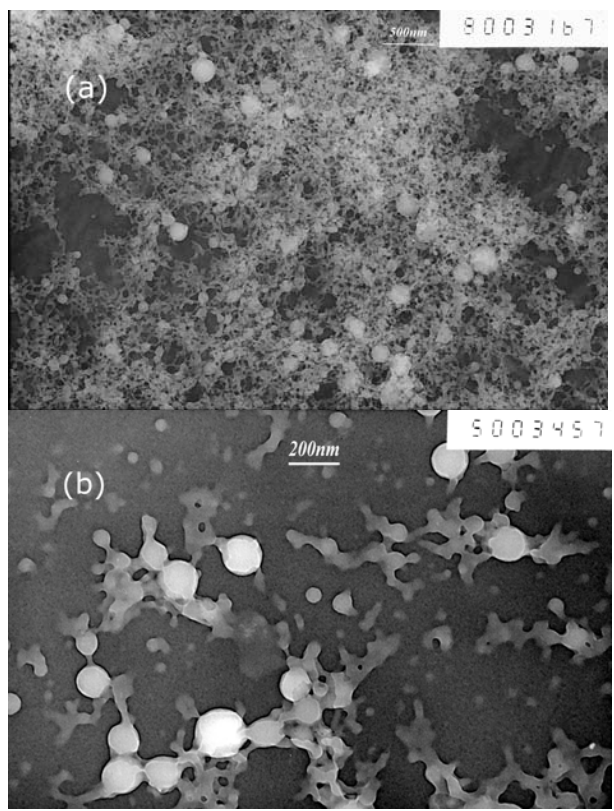

**Figure 3.** TEM morphologies of organic nanoparticles (bright spheres) dispersed in polymer matrix where the organic nanoparticles were prepared with the concentration of 4-vinyl-pyridine monomer units or MY molecules in (a) $1.0\times10^{-3}$ M and in (b) $1.0\times10^{-5}$ M, respectively.

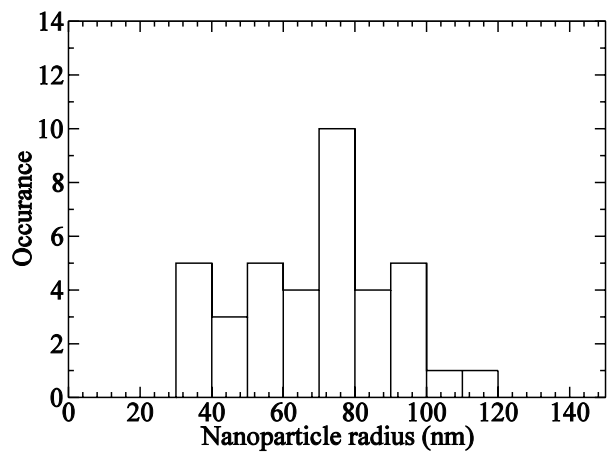

**Figure 4.** Statistical histogram of the radius of organic nanoparticles in the TEM micrographs of Figure 3.

**3.2. Strongly broadened Raman shift band**

Figure 4 shows a strongly broadened Raman shift band in the range of 1200-1700 cm$^{-1}$ for the polymer nanocomposite containing organic nanoparticles. This broad Raman shift band is not simply caused by spectral overlap of the P4VP-*r*-PAN copolymer and MY molecules because the Raman shift peaks of P4VP-r-PAN are very sharp and dicrete within this range [46]. By comparing to the experimental Raman shift spectrum of natural MY (see Figure 5), this strongly broadened Raman shift band can be related to the aggregation of MY molecules within the organic nanoparticles. As shown Figure 4, the calculated Raman activity spectrum of MY acid form is in good agreement with the experimental Raman shift spectrum of natural MY, particularly in terms of peak positions. Hence, the Raman activity calculation result can be used to understand the relationship between the measured

Raman shifts and the corresponding molecular vibrations. As indicated by the Raman activity calculation, three strong measured Raman shift peaks at 1408 cm$^{-1}$, 1439 cm$^{-1}$ and 1595 cm$^{-1}$ are caused by the N=N stretch coupled with the aromatic C-C stretch; two mediately-strong peaks at 1148 cm$^{-1}$ and 1193 cm$^{-1}$ are ascribed to the wagging of C-H on phenyl rings; and a shoulder peak at ~1310 cm$^{-1}$ are assigned to the aromatic C-C stretch on the phenyl ring. Compared to the Raman shifts of natural MY, the Raman shift band for the polymer nanocomposite containing organic nanoparticles is broadened by ~130 cm$^{-1}$ toward the lower frequency regime (from 1250 cm$^{-1}$ of natural MY) and by ~40 cm$^{-1}$ toward the higher frequency regime (from 1650 cm$^{-1}$ of natural MY). By further computational simulation, the vibrational frequencies change in a model structure consisted of two co-facially packed MY acid molecules comparing to those of a natural MY acid molecule. Thus, the close packing of MY molecules in the polymer nanocomposite containing organic nanoparticles may be a major source for the strong Raman broadening effect.

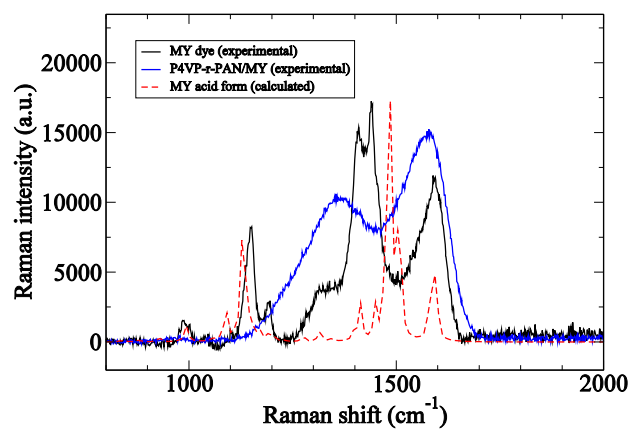

**Figure 5.** Experimental Raman shift spectra of natural MY (black solid line) and the P4VP-*r*-PAN/MY nanocomposite (blue solid line) under 632 nm excitation. The caculated Raman activitiy spectrum for MY acid form is plotted (red dash line) to compare with the experimental Raman shift spectrum of natural MY.

## 4. Conclusions

In conclusion, we have described a new class of a new class of polymer nanocomposite with organic nanoparticles dispersed in polymer matrix *via* the self-assembly of protonated PV4P-*r*-AN and amphiphilic metanil yellow dye molecules. The organic nanoparticles show to be in spherical shape with the sizes of 30-120 nm in radius. A strongly broadened Raman shift band was probed for the polymer nanocomposite containing MY dye molecules. Computational simulation suggests that the close packing of metanil yellow molecules in the polymer nanocomposite may be a major source for the strong Raman broadening effect.

**Acknowledgements**

We acknowledge financial support from the Chinese Doctorate-Granting Institution Special Research Fund (200806101010), and thank David N. Beratan and Weitao Yang at Duke University for helpful discussion.